\pdfoutput=1 % to force pdfLaTeX
\documentclass[prl,twocolumn,amssymb,10pt,aps]{revtex4-1}
\usepackage{amsmath,amssymb,mathrsfs}
\usepackage{graphicx,epsfig,float}
\usepackage{hyperref}
\usepackage{eucal}
\usepackage{bm}
\newcommand{\Real}{\mathbb{R}}
\newcommand{\Com}{\mathbb{C}}

\newcommand{\sii}{L^2}
\newcommand{\si}{L^1}

\newcommand{\dom}{\mathop{\mathsf{dom}}}
\newcommand{\Hilbert}{\mathcal{H}}
\newcommand{\eps}{\varepsilon}
\newcommand{\der}{\mathrm{d}}
\usepackage{color}
\usepackage[normalem]{ulem}
\definecolor{DarkGreen}{rgb}{0,0.5,0.1} % David

\newcommand\soutD{\bgroup\markoverwith
{\textcolor{DarkGreen}{\rule[.5ex]{2pt}{1pt}}}\ULon}
\newcommand\soutP{\bgroup\markoverwith
{\textcolor{blue}{\rule[.5ex]{2pt}{1pt}}}\ULon}
\newcommand{\Hm}[1]{\leavevmode{\marginpar{\tiny%
$\hbox to 0mm{\hspace*{-0.5mm}$\leftarrow$\hss}%
\vcenter{\vrule depth 0.1mm height 0.1mm width \the\marginparwidth}%
\hbox to
0mm{\hss$\rightarrow$\hspace*{-0.5mm}}$\\\relax\raggedright #1}}}

\begin{document}
\title{Bound states in semi-Dirac semi-metals}
\author{D.~Krej\v{c}i\v{r}\'{i}k$^1$}
\email{David.Krejcirik@fjfi.cvut.cz}
\author{P.R.S.~Antunes$^2$}
\affiliation{$^1$Department of Mathematics, 
Faculty of Nuclear Sciences and Physical Engineering, 
Czech Technical University in Prague, 
Trojanova 13, 12000 Prague 2, Czech Republic
\\
$^2$ Department of Science and Technology, Universidade Aberta and Group of Mathematical Physics, FCUL,
Campo Grande, Edif\'icio 6, Piso 1,
1749-016 Lisbon, Portugal}
\date{3 July 2020}

\begin{abstract}
New insights into transport properties of nanostructures
with a linear dispersion along one direction 
and a quadratic dispersion along another
are obtained by analysing their spectral stability properties
under small perturbations.
Physically relevant sufficient and necessary conditions 
to guarantee the existence of discrete eigenvalues
are derived under rather general assumptions on external fields.
One of the most interesting features of the analysis is 
the evident spectral instability of the systems in the weakly coupled regime.
The rigorous theoretical results are illustrated by numerical experiments
and predictions for physical experiments are made.
\end{abstract}

\maketitle

%------%
% BODY %
%------%

Semi-Dirac semi-metals have attracted a lot of attention in the last decade; 
see, e.g., \cite{Pardo-Pickett_2009,Banerjee_2009,Delplace-Montambaux_2010,
Saha_2016,Banerjee-Narayan_2020}
and references therein.
The most striking feature of these recently discovered nanostructures
is that they exhibit unprecedented band structure properties:
(electron or hole) quasiparticles disperse linearly in one direction 
and quadratically in the orthogonal direction.
The situation is neither conventional zero-gap semiconductor-like, 
nor graphene-like, but has in some sense aspects of both.

Using a tight-binding model of spinless fermions,
it is commonly accepted that the Hamiltonian
\begin{equation}
  H_0 := 
  \begin{pmatrix}
    -i\partial_y & -\partial_x^2 + \delta \\
    -\partial_x^2 + \delta & i\partial_y
  \end{pmatrix}
\end{equation}
is the right low-energy description of the unperturbed system.
Here we disregard all the physical constants of 
\cite{Banerjee_2009,Delplace-Montambaux_2010},
for they can always be considered to be equal to~$1$ 
by suitably re-scaling the space variables 
$\bm{r} := (x,y) \in \Real^2$,
except for the gap parameter~$\delta$
which we assume to be a \emph{positive} constant. 

We understand~$H_0$ as the operator acting in
the Hilbert space $\Hilbert := \sii(\Real^2)^2$
consisting of all $\Com^2$-valued functions
$$
  \psi = 
  \begin{pmatrix}
    \psi_1 \\ \psi_2
  \end{pmatrix}
  \qquad \mbox{such that} \qquad
  \|\psi\|_{\Hilbert}^2 := \int_{\Real^2} |\psi|^2 < \infty 
  \,,
$$
where $|\psi|:=\sqrt{|\psi_1|^2+|\psi_2|^2}$
is the usual Euclidean norm
and $\sii(\Real^2)$ is the Lebesgue space
of square-integrable functions over~$\Real^2$.
For the operator domain, we take
$$
  \dom H_0 := \left\{
    \psi \in \Hilbert : \
    \partial_x\psi, \, \partial_x^2\psi, \, \partial_y\psi \in \Hilbert
  \right\}
  ,
$$
which, in contrast to the conventional Dirac operator,
is a \emph{proper} subset of the Sobolev space $H^1(\Real^2)^2$.
Anyway, applying the Fourier transform in the spirit of
\cite[\S\,V.5.4]{Kato} or \cite[\S\,1.4]{Thaller},
%which consists of applying the Fourier transform~$\mathcal{F}$ 
%and diagonalising the transformed matrix multiplication operator 
%$\mathcal{F}H_0\mathcal{F}^{-1}$, it is easily verified that~$H_0$
%is unitarily equivalent to
%
%\begin{equation}
%  \hat{H}_0 := 
%  \begin{pmatrix}
%    \sqrt{(X^2+\delta)^2+Y^2} & 0 \\
%    0 & -\sqrt{(X^2+\delta)^2+Y^2}
%  \end{pmatrix}
%\end{equation}
%
%with $\dom \hat{H}_0 := \dom H_0$,
%where $X,Y \in \Real^2$ are the dual variables of~$\mathcal{F}$.
it is easily verified that~$H_0$ is self-adjoint 
and that its spectrum is given by
\begin{equation}\label{spectrum}
  \sigma(H_0) = (-\infty,-\delta] \cup [\delta,\infty)
  \,.
\end{equation}
Moreover, the total spectrum is purely absolutely continuous,
which is traditionally interpreted 
(see~\cite{Bruneau-Jaksic-Last-Pillet_2016} for a nice overview)
as the existence of transport
for the whole set of energies~$E$ satisfying $|E| \geq \delta$.

In this paper, we are concerned with spectral stability properties of~$H_0$.
More specifically, we consider a general matrix multiplication operator
\begin{equation}\label{potential}
  V := 
  \begin{pmatrix}
    V_{11} & V_{12} \\
    V_{21} & V_{22}
  \end{pmatrix}
  ,
\end{equation}
whose coefficients are bounded complex-valued functions
$V_{11},V_{12},V_{21},V_{22}: \Real^2 \to \Com$,
and study the spectrum of the perturbed operator 
$$
  H_\eps := H_0 + \eps V
  \,, \qquad
  \dom H_\eps = \dom H_0
  \,,
$$ 
as the positive coupling parameter~$\eps$ tends to zero.
To make~$H_\eps$ self-adjoint, we always assume 
that~$V_{11}$ and~$V_{22}$ are in fact real-valued,
while~$V_{12}$ and~$V_{21}$ are allowed to be complex-valued
but the Hermiticity relation $V_{21}=\overline{V_{12}}$ is postulated.
In addition, we assume that $V_{11},V_{12},V_{22}$ are vanishing at infinity, 
in order to have (cf.~\cite[\S\,4.3.4]{Thaller})
the stability of the essential spectrum
%$\sigma_\mathrm{ess}(H_\eps) = \sigma(H_0)$.
%
\begin{equation}\label{essential}
  \sigma_\mathrm{ess}(H_\eps) = (-\infty,-\delta] \cup [\delta,\infty)
  \,.
\end{equation}

Recall that the essential spectrum is composed 
of accumulation points of the spectrum
and possibly also of infinitely degenerate eigenvalues.
For the stability issues, 
we are more interested in the discrete spectrum $\sigma_\mathrm{disc}(H_\eps)$,
which consists of isolated eigenvalues of finite multiplicities
in the essential spectral gap $(-\delta,\delta)$.
Physically, the eigenvalues are energies of bound states of~$H_\eps$
representing stationary solutions of the time-dependent Dirac equation. 
Our objective is to derive physically relevant sufficient 
and necessary conditions for the existence of the discrete eigenvalues.
Contrary to the Schr\"odinger case, this is methodologically 
by no means evident, for no direct variational principles are available
for the operator~$H_\eps$ due to its unboundedness from below. 

Our strategy to overcome this difficulty is to pass to the square $H_\eps^2$,
which is a non-negative operator, 
apply the standard variational principle 
(see, e.g., \cite[\S\,4.5]{Davies})
to it
and employ the spectral mapping equivalence
\begin{equation}\label{mapping}
  E \in \sigma(H_\eps) 
  \quad\Longleftrightarrow\quad 
  E^2 \in \sigma(H_\eps^2)
\end{equation}
valid for all real energies~$E$.
Consequently, in order to ensure that there exists a discrete eigenvalue
$E \in (-\delta,\delta)$, it is enough to construct a test function
$\psi \in \dom H_0$ such that 
\begin{equation}
  Q_\eps[\psi] := 
  \|H_\eps\psi\|_{\Hilbert}^2 - \delta^2 \, \|\psi\|_{\Hilbert}^2 < 0 \,.
 \end{equation}

Motivated by the theory of quantum waveguides~\cite{DEK2},
we choose the test function as follows.
Observing that, formally(!), 
$H_0^2\psi^\pm \stackrel{!}{=} \delta^2\psi^\pm$,
where 
\begin{equation}\label{inadmissible}
  \psi^+ := \begin{pmatrix} 1 \\ 0\end{pmatrix}
  \qquad\mbox{and}\qquad
  \psi^- := \begin{pmatrix} 0 \\ 1\end{pmatrix}
  ,
\end{equation}
we see that~$\psi^\pm$ are generalised eigenvectors of~$H_0^2$
corresponding to the ionisation energy~$\delta^2$.
Therefore they are generalised minimisers of the functional~$Q_0$
and it is admissible to expect them to be suitable building blocks 
for possible minimisers of~$Q_\eps$ as well, 
at least if~$\eps$ is small.  
Still formally(!), one easily computes
\begin{equation}\label{formal}
\begin{aligned}
  Q_\eps[\psi^+] 
  &\!\stackrel{!}{=} \!\! 
  \int_{\Real^2} \!(\eps^2 |V_{11}|^2 + \eps^2 |V_{12}|^2 + 2\delta \eps \, \Re V_{12})
  =: I_\eps^+
  ,
  \\
  Q_\eps[\psi^-] 
  &\!\stackrel{!}{=} \!\!
  \int_{\Real^2} \!(\eps^2 |V_{22}|^2 + \eps^2 |V_{12}|^2 + 2\delta \eps \, \Re V_{12})
  =: I_\eps^-
  .
\end{aligned}
\end{equation}
To make sense of the integrals, 
we henceforth assume $V_{11}, V_{22} \in \sii(\Real^2)$
and $V_{12} \in \sii(\Real^2) \cap \si(\Real^2)$.
We have thus obtained the following sufficient condition:
\begin{equation}\label{sufficient}
  \left( \, I_\eps^+ < 0 \quad \mbox{or} \quad I_\eps^- < 0 \, \right)
  \quad\Longrightarrow\quad
  \sigma_\mathrm{disc}(H_\eps) \not= \varnothing
  \,,
\end{equation}
meaning that~$H_\eps$ possesses at least one isolated eigenvalue 
of finite multiplicity located in the interval $(-\delta,\delta)$. 
As a matter of fact, the variational principle implies 
that~$H_\eps$ possesses at least \emph{two} discrete eigenvalues
(counting multiplicities) provided that 
$I_\eps^+ < 0$ \emph{and} $I_\eps^- < 0$ hold,
because the test functions~$\psi^\pm$ are mutually orthogonal.

To justify the formal 
computations above ($\psi^\pm \not\in \Hilbert$\,!), 
we replace the inadmissible test functions~\eqref{inadmissible} 
by their regularised versions 
$\psi_n^+ := \phi_n \, \psi^\pm$ with $n>1$.
%($n>1$)
%
%\begin{equation}\label{admissible}
%  \psi_n^+ := \begin{pmatrix} \phi_n \\ 0\end{pmatrix}
%  \qquad\mbox{and}\qquad
%  \psi_n^- := \begin{pmatrix} 0 \\ \phi_n \end{pmatrix}
%  .
%\end{equation}
%
Here $\phi_n:\Real^2\to\Real$ is a smooth function of compact support 
such that $\phi_n=1$ on the disk of radius~$n$,
$\phi_n=0$ outside the disk of radius~$n^2$ 
and $\phi_n(\bm{r}) := \xi(f(r))$ elsewhere,
where
$
  f(r) := \log_n(n^2/r)
$
with $r:=|\bm{r}|$
and $\xi:\Real\to[0,1]$ is any smooth function such that  
$\xi=0$ in a right neighbourhood of~$0$
and $\xi=1$ in a left neighbourhood of~$1$.
Then the formal results~\eqref{formal} are indeed justified 
%with help of the dominated convergence 
through the limits
$Q_\eps[\psi_n^\pm] \to I_\eps^\pm$ as $n \to \infty$.
Consequently, assuming $I_\eps^+ < 0$ (respectively, $I_\eps^- < 0$),
then there exists a positive number~$n_0$ such that 
$Q_\eps[\psi_n^+] < 0$ (respectively, $Q_\eps[\psi_n^-] < 0$)
for all $n > n_0$.
Hence~\eqref{sufficient} holds true
as well as the remark about the existence of two discrete eigenvalues.

It is remarkable that the sufficient condition~\eqref{sufficient}
is always satisfied in the weakly coupled regime provided that 
\begin{equation}\label{sufficient.weak}
  \Re V_{12} < 0 
  \,.
\end{equation}
Indeed, under this condition, 
there obviously exists a positive number~$\eps_0$
such $I_\eps^+ < 0$ and $I_\eps^- < 0$ 
for all $\eps < \eps_0$.
It follows that, for all sufficiently small~$\eps$,
$H_\eps$~possesses at least two isolated eigenvalues 
of finite multiplicities located in the interval $(-\delta,\delta)$.
We interpret the result as 
the \emph{spectral instability} (or \emph{criticality}) of~$H_0$,
for there always exists an electromagnetic potential~$V$
such that the spectrum of~$H_\eps$ with an arbitrarily small~$\eps$
differs from that of~$H_0$ given by~\eqref{spectrum}.  

A special situation in which the discrete spectrum exists 
is the potential~$V$ with vanishing diagonal components $V_{11}=0=V_{22}$
and the off-diagonal component~$V_{12}$ satisfying~\eqref{sufficient.weak}.
In this case the critical coupling constant satisfies
%$
%  \eps_0 \geq 
%  -\int_{\Real^2} \Re V_{12}
%  / \int_{\Real^2} |V_{12}|^2 
%$.
%
\begin{equation}\label{coupling}
  \eps_0 \geq 
  \frac{-2\delta \, \langle \Re V_{12} \rangle}
  {\|V_{12}\|^2} 
  \,,
\end{equation}
where we abbreviate 
$\langle \Re V_{12} \rangle := \int_{\Real^2} \Re V_{12}$ 
and~$\|\cdot\|$ denotes the norm of $\sii(\Real^2)$.

At least in this special setting and if~$V_{12}$ is real-valued,
it is worth noticing that~\eqref{sufficient.weak} represents also
a necessary condition for the existence of discrete spectrum. 
To see it, let us now assume that $V_{11}=0=V_{22}$
and 
\begin{equation}\label{necessary}
  V_{12} = V_{21} \geq 0 
  \,.
\end{equation}
From the first component of the eigenvalue equation $H_0\psi = E\psi$,
we get 
$
  \psi_2 = -R \, (-i\partial_y - E) \psi_1
$,
where the inverse 
$R := (-\partial_x^2+\delta+\eps V_{12})^{-1}$
is a well defined isomorphism on $\sii(\Real^2)$
because of~\eqref{necessary}.
Plugging this relationship between~$\psi_1$ and~$\psi_2$ 
into the second component of the eigenvalue equation,
we arrive at the functional identity
\begin{equation*}
  (-\partial_x^2+\delta+\eps V_{21})\psi_1
  - (i\partial_y-E) R (-i\partial_y-E)\psi_1 = 0
  \,.
\end{equation*}
Multiplying both sides by~$\overline{\psi_1}$, 
integrating over~$\Real^2$,
taking the real part of the obtained scalar identity
and using the self-adjointness of~$R$, 
we get 
\begin{multline}\label{scalar}
  \|\partial_x\psi_1\|^2 + \delta \, \|\psi_1\|^2 + (\psi_1,\eps V_{21}\psi_1)
  + \|R^{1/2}\partial_y\psi_1\|^2	
  \\
  = E^2 \|R^{1/2}\psi_1\|^2	
  \,,
\end{multline}
where $(\cdot,\cdot)$ denotes
the inner product of $\sii(\Real^2)$
associated with~$\|\cdot\|$.
Since~$V_{12}$ is assumed to be real-valued,
$-\partial_x^2+\delta+\eps V_{12}(x,y)$ considered 
as an operator in $\sii(\Real)$ parametrically dependent on~$y$
is self-adjoint.
Recalling in addition that $V_{12} \geq 0$ vanishes at infinity, 
so that the spectrum of the one-dimensional 
Schr\"odinger operator equals $[\delta,\infty)$,
one has the estimate
$$
  \|R^{1/2}\psi_1\|^2 \leq \|R\| \|\psi_1\|^2
  = \delta^{-1}\|\psi_1\|^2 \,.
$$
Using this bound in~\eqref{scalar},
we finally get $\delta^2 \leq E^2$,
which proves that the discrete spectrum of~$H_\eps$
is empty in view of~\eqref{mapping} and~\eqref{essential}.

Our last theoretical objective is to establish quantitative bounds
for the discrete eigenvalues existing under the hypothesis~\eqref{sufficient.weak}
in the weakly coupled regime. 
To this aim, we henceforth assume that
the bounded functions $V_{11}, V_{12}, V_{22}$ are compactly supported.
As in the beginning, we allow~$V_{12}$ to be complex-valued. 
By the variational principle, one has the bound
$$
  E^2 - \delta^2 \leq \frac{Q_\eps[\psi_n^\pm]}{\|\psi_n^\pm\|_\Hilbert^2}
  \,,
$$
where the test functions~$\psi_n^\pm$ are 
the regularised versions of~\eqref{inadmissible} as above.
%given by~\eqref{admissible}. 

Let us begin with the test function~$\psi_n^+$.
One has
$$
\begin{aligned}
\lefteqn{  
  \|H_\eps\psi_n^+\|_\Hilbert^2 
}\\
  &= \|(-\partial_x^2+\delta+\eps V_{21})\phi_n\|^2
  + \|(-i\partial_y+\eps V_{11})\phi_n\|^2 
  \\
  &= 
  %\|\partial_x^2\phi_n\|^2 + \delta^2 \|\phi_n\|^2 + \eps^2 \|V_{12}\|^2
  %+ 2 \delta \|\partial_x\phi_n\|^2
  %\\
  %& \quad
  %+ 2\delta \, \eps \, \langle \Re V_{12} \rangle
  %+ \|\partial_y\phi_n\|^2 + \eps^2 \|V_{11}\|^2
  \|\partial_x^2\phi_n\|^2 + \delta^2 \|\phi_n\|^2
  + 2 \delta \|\partial_x\phi_n\|^2 + \|\partial_y\phi_n\|^2
  + I_\eps^+
  \,,
\end{aligned}
$$
where the second equality holds for all sufficiently large~$n$
when $V_{12}$ and $\partial_x^2\phi_n$ (and $V_{11}$ and $\partial_y\phi_n$)
have disjoint supports.
Using the chain rule when differentiating~$\phi_n$,
estimating the derivative of~$\xi$ by its maximal value
$\|\xi'\|_\infty:=\max_{[0,1]} |\xi'|$
and passing to polar coordinates, we have
$$
  \|\partial_x\phi_n\|^2 
  \leq \frac{\|\xi'\|_\infty^2}{\log^2n}
  %\left\|\frac{x}{r^2}\right\|^2
  \int_{\{n < r < n^2\}} \frac{x^2}{r^4} \, \der x \, \der y
  %= \frac{\pi \|\xi'\|_\infty^2}{\log n}
  = \frac{c_1}{\log n}
  \,,
$$
where $c_1 := \pi \|\xi'\|_\infty^2$.
The same estimate holds for $\|\partial_y\phi_n\|$.
Similarly,
$$
\begin{aligned}
  \|\partial_x^2\phi_n\|^2 
  &\leq \frac{2\,\|\xi''\|_\infty^2}{\log^4n} 
  \int_{\{n < r < n^2\}} \frac{x^4}{r^8} \, \der x \, \der y
  %\left\|\frac{x^2}{r^4}\right\|^2
  \\
  &\quad + \frac{2\,\|\xi'\|_\infty^2}{\log^2n} 
  \int_{\{n < r < n^2\}} \frac{(x^2-y^2)^2}{r^8} \, \der x \, \der y
  %\left\|\frac{x^2-y^2}{r^4}\right\|^2 
  \\
  &= 
  \left( \frac{3\pi \|\xi''\|_\infty^2}{4 \log^4n}
  + \frac{\pi \|\xi'\|_\infty^2}{\log^2n} \right)
  \left(\frac{1}{n^2}-\frac{1}{n^4}\right)
  \\
  &\leq \left( \frac{3\pi \|\xi''\|_\infty^2}{4\log n}
  + \frac{\pi \|\xi'\|_\infty^2}{\log n} \right) \frac{1}{e^2}
  %< \frac{\|\xi''\|_\infty^2+\|\xi'\|_\infty^2}{\log^2n}
  =: \frac{c_2}{\log n}
  \,,
\end{aligned}
$$
where~$e$ is the base of the natural logarithm~$\log$
and the last, crude estimate holds for all $n \geq e$.

Using these estimates, we observe that 
$Q_\eps[\psi_n^+] \to I_\eps^+$ as $n \to \infty$,
in agreement with our claim above. 
Under the hypothesis~\eqref{sufficient.weak},
the limit~$I_\eps^+$ is negative for all sufficiently small~$\eps$;
in fact, whenever
$$
  \eps < \frac{-2\delta\, \langle \Re V_{12}\rangle}
  {\|V_{11}\|^2+\|V_{12}\|^2}
  \,.
$$
Henceforth we therefore assume this inequality
and then choose $n \geq e$ so large 
that $Q_\eps[\psi_n^+]$ is negative. 
Finally, using
$$
  \|\psi_n^+\|_\Hilbert^2
  = \|\phi_n\|^2 \leq \int_{\{r < n^2\}} 1 \, \der x \, \der y
  = \pi n^4 \,,
$$
it follows that 
\begin{equation*}
  E^2 - \delta^2 \leq \frac{1}{\pi n^4} \left(
  \frac{c}{\log n} + I_\eps^+
  \right) 
  =: g^+(\eps,n)
  \,,
\end{equation*}
where $c := c_1+2\delta c_1 + c_2$.

Using the test function~$\psi_n^-$ instead of~$\psi_n^+$,
the proof follows analogously. In fact, it is enough 
to replace~$V_{11}$ by~$V_{22}$ (and thus~$I_\eps^+$ by~$I_\eps^-$)  
in the formulae above.
In particular, we have $E^2 - \delta^2 \leq g^-(\eps,n)$,
where~$g^-$ is defined as~$g^+$ 
with~$I_\eps^+$ being replaced by~$I_\eps^-$. 

The function $n \mapsto g^\pm(\eps,n)$ achieves its negative minimum 
for the critical value~$n_\eps^\pm$ satisfying 
$$
  \frac{1}{\log n_\eps^\pm} 
  %= \frac{2}{c} \left(-c+\sqrt{c^2-c I_\eps^+}\right)
  := \frac{-2\,I_\eps^\pm}{c+\sqrt{c^2-c I_\eps^\pm}}
$$
(notice that $n_\eps^\pm \to \infty$ as $\eps \to 0$). 
In summary, we have got an explicit quantitative bound
for the discrete energies
\begin{equation}\label{bound}
  E^2 - \delta^2 \leq g^\pm(\eps,n_\eps^\pm)
  \,.
\end{equation}
In the weakly coupled regime, one has
\begin{equation}\label{bound.weak}
  g^\pm(\eps,n_\eps^\pm) \approx 
  -\frac{\delta^2 \, \langle \Re V_{12}\rangle^2 \, \eps^2}{\pi c} \
  \exp\left(\frac{2c}{\delta \, \langle \Re V_{12}\rangle \, \eps}\right)
\end{equation}
as $\eps \to 0$.

Now we turn to numerical verifications of the established theoretical results.
Our numerical scheme consists in expanding the components~$\psi_1,\psi_2$
of an eigenvector $\psi \in \dom H_0 \subset \Hilbert$ of~$H_\eps$
corresponding to an eigenvalue~$E$ 
into a basis $\{\varphi_j\}_{j=1}^\infty$ of~$\sii(\Real^2)$:
$$
  \psi_1 = \sum_{j=1}^\infty a_j \, \varphi_j
  \qquad\mbox{and}\qquad
  \psi_2 = \sum_{j=1}^\infty b_j \, \varphi_j
  \,,
$$ 
where $a_j := (\varphi_j,\psi_1)$ and $b_j := (\varphi_j,\psi_2)$.
The eigenvalue problem $H_\eps\psi=E\psi$ in~$\Hilbert$
is cast into a system of algebraic equations 
for the coefficients 
$\bm{a} := \{a_j\}_{j=1}^\infty$  
and $\bm{b} := \{b_j\}_{j=1}^\infty$ 
in the sequence space~$\ell^2$: 
$$
  \begin{pmatrix}
    \bm{C_{11}} & \bm{C_{12}} \\
    \bm{C_{21}} & \bm{C_{22}}
  \end{pmatrix}
  \begin{pmatrix}
    \bm{a} \\ \bm{b}
  \end{pmatrix}
  = E   \begin{pmatrix}
    \bm{D} & \bm{0} \\
    \bm{0} & \bm{D}
  \end{pmatrix}
  \begin{pmatrix}
    \bm{a} \\ \bm{b}
  \end{pmatrix}
  \,,
$$
where
$$
\begin{aligned}
  \bm{C_{11}} &:= \big\{
  (\varphi_k,-i\partial_y\varphi_j)
  + (\varphi_k,\eps V_{11}\varphi_j)
  \big\}_{k,j=1}^\infty
  \,, 
  \\
  \bm{C_{12}} &:= \big\{
  (\varphi_k,(-\partial_x^2+\delta)\varphi_j)
  + (\varphi_k,\eps V_{12}\varphi_j)
  \big\}_{k,j=1}^\infty
  \,,
  \\
  \bm{C_{21}} &:= \big\{
  (\varphi_k,(-\partial_x^2+\delta)\varphi_j)
  + (\varphi_k,\eps V_{21}\varphi_j)
  \big\}_{k,j=1}^\infty
  \,, 
  \\
  \bm{C_{22}} &:= \big\{
  (\varphi_k,i\partial_y\varphi_j)
  + (\varphi_k,\eps V_{22}\varphi_j)
  \big\}_{k,j=1}^\infty
  \,, 
  \\
  \bm{D} &:=  \big\{
  (\varphi_k,\varphi_j)
  \big\}_{k,j=1}^\infty
  \,. 
\end{aligned}
$$
The numerical approximation consists in replacing the infinite
matrices by finite ones. The obtained system can be then solved
by standard tools of numerical linear algebra.
Since no natural basis seems to be available for the problem,
we choose the basis consisting of Gaussian radial basis function centered at a set of scattered nodes, in the line of the Radial Basis Function Method.

In our numerical experiments, we considered potentials~$V$ with coefficients
being either piecewise-constant or fastly decaying functions. 
In both cases, we got the same qualitative behaviour of the eigenvalues
and a quantitative verification of the spectral enclosure~\eqref{bound}. 
Therefore it is expected that this bound is more universal. %though probably not very sharp.

\begin{figure}[h]
\centering 
\includegraphics[width=0.5\textwidth]{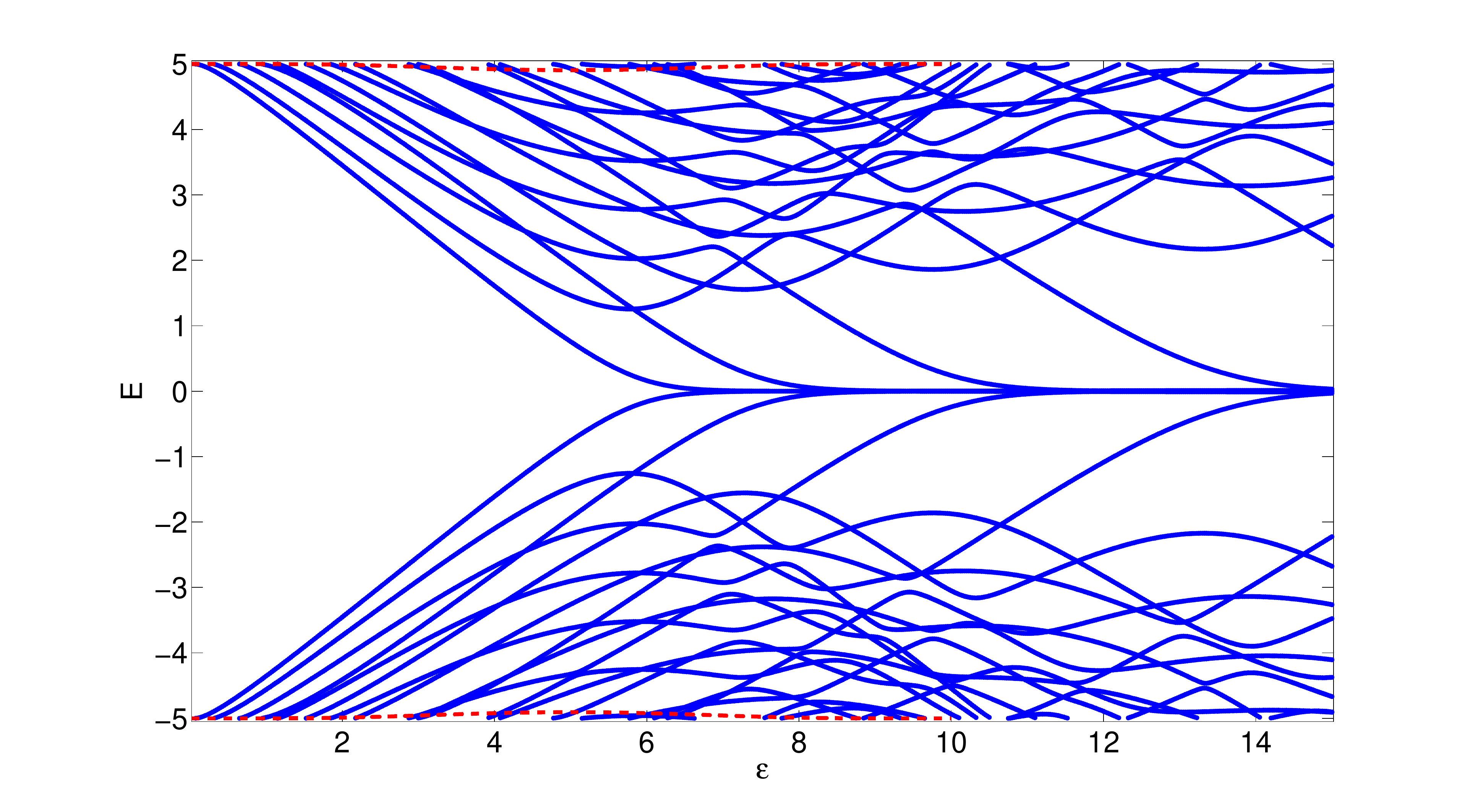}%
\\
\includegraphics[width=0.5\textwidth]{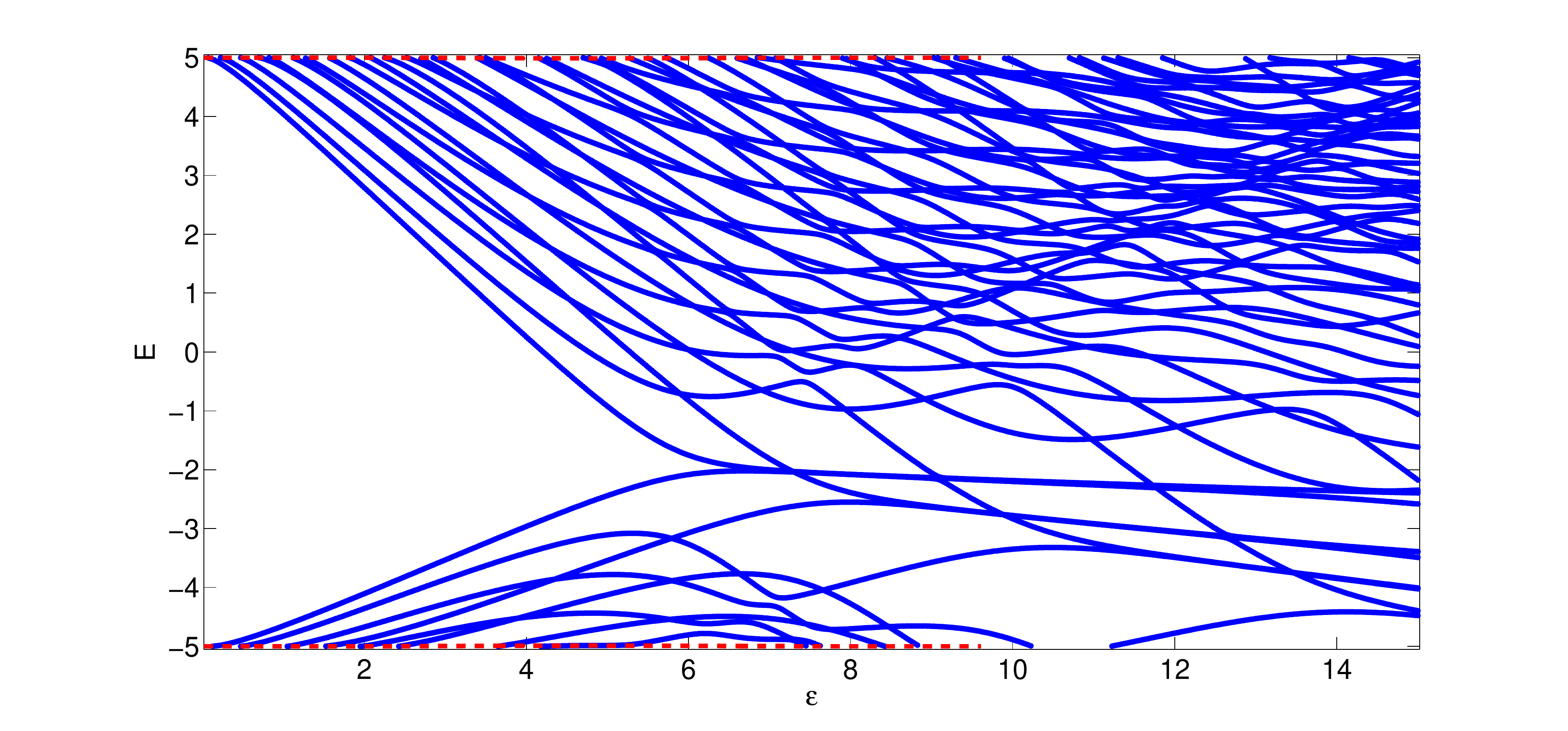}%
\caption{Plots of eigencurves $E(\eps)$ (in blue) 
and the bounds~$h(\eps)$ of \eqref{boundeig} (in red) for $\delta=5$.
The apparently symmetric setting in the upper figure
is due to the choice $V_{11}=0=V_{22}$ and $V_{21}=-\chi_D$, 
while the lower figure corresponds to
$V_{21}=-\chi_D,$ $V_{11}=0.2\chi_D,$ $V_{22}=-0.9\chi_D$.} \label{Fig.both}
\end{figure}

The dependence of several eigenvalues (blue curves)
on the coupling parameter~$\eps$ in the gap $(-\delta,\delta)$
is depicted in Figure~\ref{Fig.both} for two seetings.
In both cases, $\chi_D$~denotes the characteristic function 
of the disk~$D$ of radius~$2$ centered at the origin and $\delta=5$.
We also plot the bounds~$\pm h$ (red curves) of the estimates
\begin{equation}\label{boundeig}
  -h(\eps) \leq E(\eps)\leq h(\eps) := \sqrt{ \delta^2  +g^\pm(\eps,n_\eps^\pm)}
  \,
\end{equation}
directly obtained from \eqref{bound}.
It turns out that the bounds~\eqref{boundeig}
become too crude for larger values of~$\eps$.

\begin{figure}[ht]
\centering 
\includegraphics[width=0.45\textwidth]{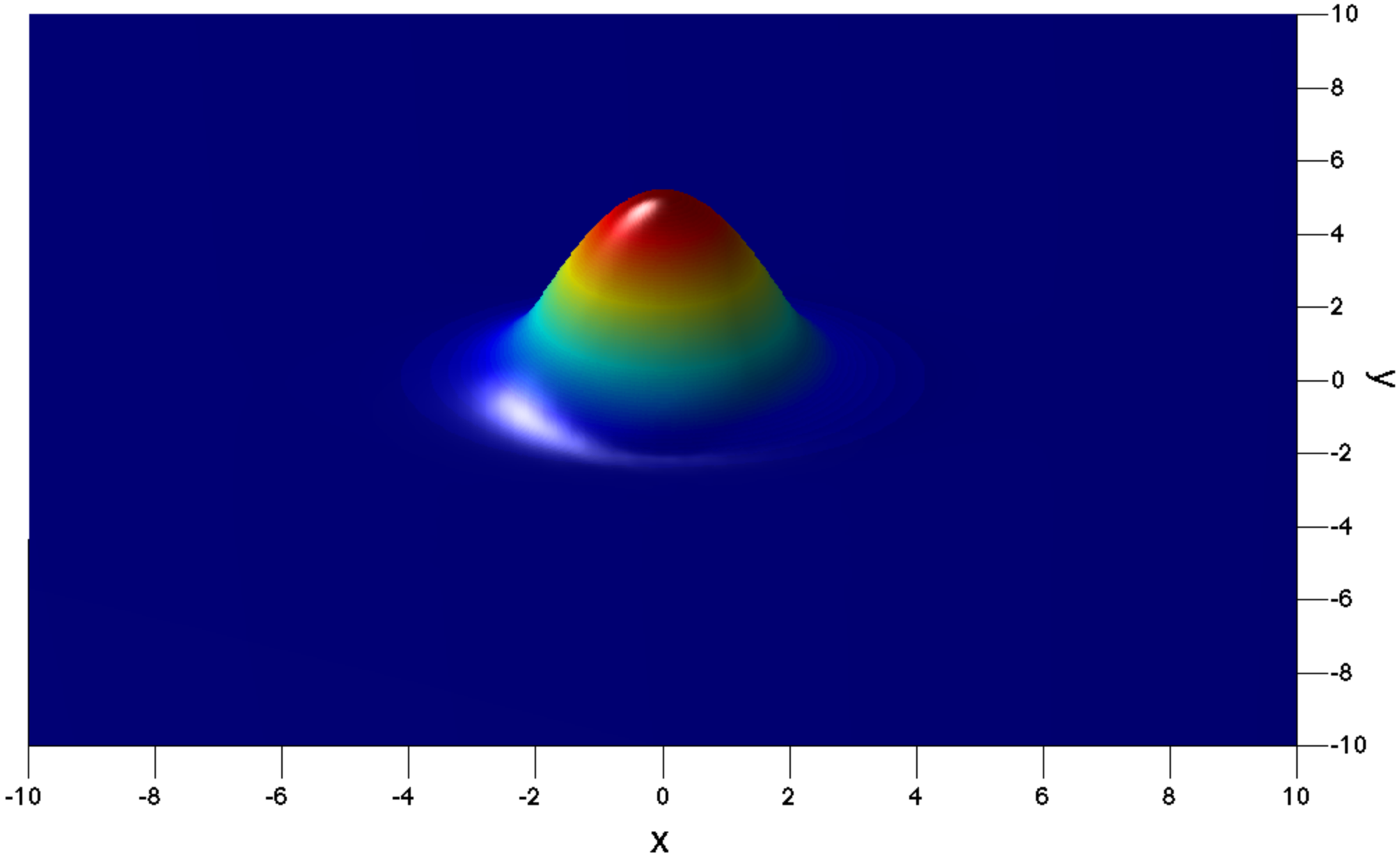}
\\
\includegraphics[width=0.45\textwidth]{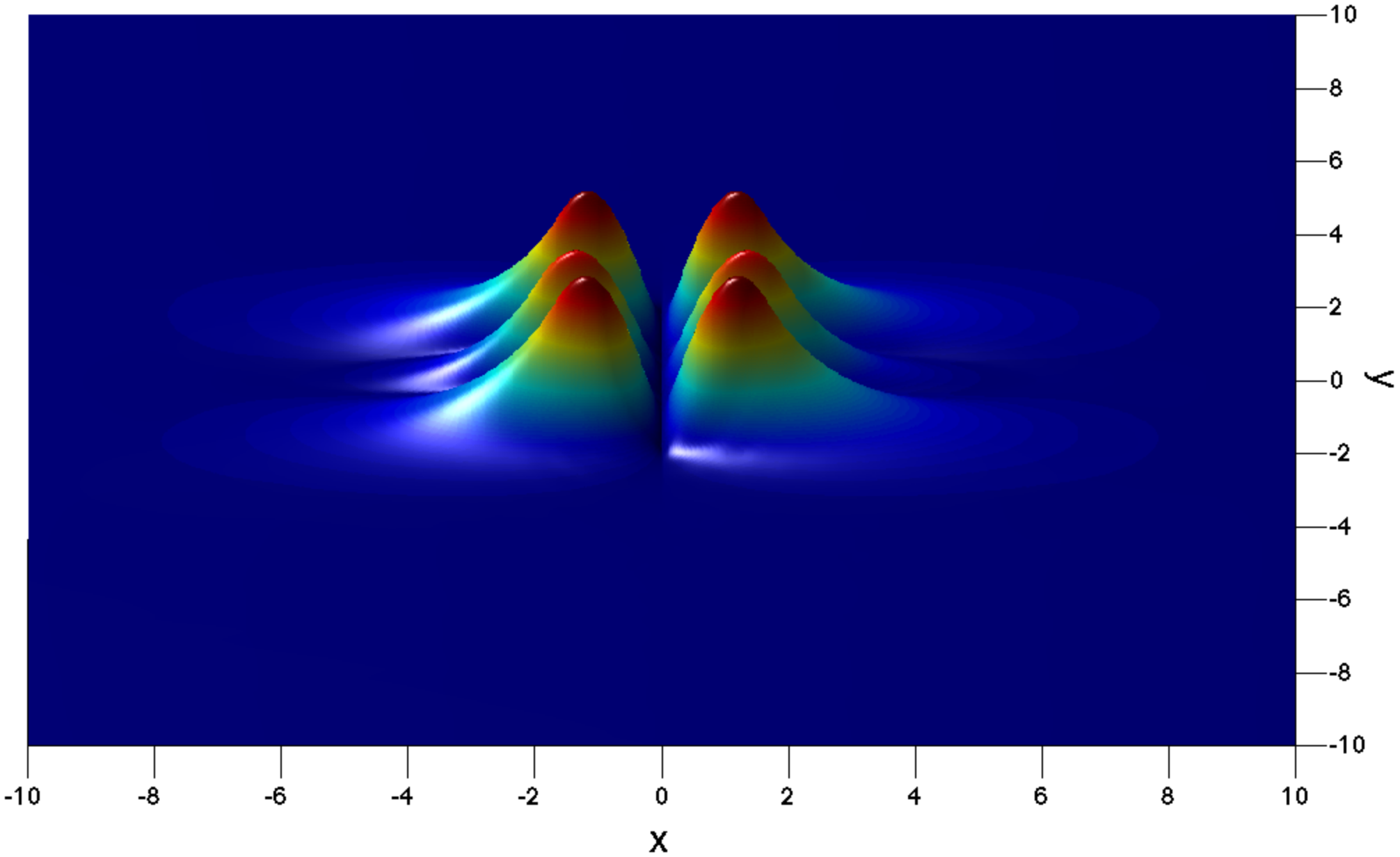} 
\caption{Plots of the magnitude~$|\psi|$ of eigenfunctions~$\psi$ corresponding to eigenvalues $E\approx 2.9893$ (up) and $E\approx 4.8284$ (down)
of the symmetric setting of Figure~\ref{Fig.both}
for $\eps=2.5$.}\label{fig:eigenfunctions}
\end{figure}

Figure~\ref{fig:eigenfunctions} visualises the ground and excited states.

In conclusion, we have derived sufficient and necessary conditions 
for the existence of discrete energies in semi-Dirac semi-metals
perturbed by general local electromagnetic fields.
The existence of bound states is particularly ensured in 
the regime of weak coupling provided that the off-diagonal
component of the perturbation is attractive 
in the sense of~\eqref{sufficient.weak}.
On the other hand, the discrete spectrum is empty in the opposite regime 
of real-valued repulsive off-diagonal component
and absent diagonal components.
We have also derived an explicit quantitative bound~\eqref{bound} 
for the discrete energies.
Numerical experiments support our theoretical results
and predict the existence of excited states as well.

Because of the tremendous progress in manipulation with
materials whose low-energy excitations are described by semi-Dirac fermions,
it is our belief that an experimental verification of our theoretical
predictions is within the reach of contemporary physics.
The simplest experimental setting should be considering
an electromagnetic potential~\eqref{potential} 
with $V_{11}=0=V_{22}$ and~$V_{12}=\overline{V_{21}}$
being a locally distributed perturbation
(possibly piecewise constant).	
We predict that the transport properties of the material    
should significantly depend on the sign of~$\Re V_{12}$.
Is the estimate~\eqref{coupling} on the critical coupling sharp?
Do the bound state energies follow the theoretical estimate~\eqref{bound}
with~\eqref{bound.weak} in the weakly coupled regime?

The present model is challenging also from purely mathematical perspectives.
Because of unavailability of an explicit form of the kernel
of the resolvent operator of the unperturbed Hamiltonian~$H_0$, 
we have not been able 
to apply the traditional approach to weakly coupled bound states
based on the Birman--Schwinger principle
(see the classical reference~\cite{Si} in the Schr\"odinger case). 
In particular, we leave as an open problem how to establish 
a (good) lower bound for discrete energies complementing~\eqref{bound},
without speaking about the exact asymptotics as $\eps \to 0$.   
It is also challenging to study perturbations of the non-self-adjoint
model recently introduced in~\cite{Banerjee-Narayan_2020}.

This project was partially supported by GA\v{C}R grant No.~20-17749X.

\bigskip
%\vfill
%
%

\begin{thebibliography}{11}%
\makeatletter
\providecommand \@ifxundefined [1]{%
 \@ifx{#1\undefined}
}%
\providecommand \@ifnum [1]{%
 \ifnum #1\expandafter \@firstoftwo
 \else \expandafter \@secondoftwo
 \fi
}%
\providecommand \@ifx [1]{%
 \ifx #1\expandafter \@firstoftwo
 \else \expandafter \@secondoftwo
 \fi
}%
\providecommand \natexlab [1]{#1}%
\providecommand \enquote  [1]{``#1''}%
\providecommand \bibnamefont  [1]{#1}%
\providecommand \bibfnamefont [1]{#1}%
\providecommand \citenamefont [1]{#1}%
\providecommand \href@noop [0]{\@secondoftwo}%
\providecommand \href [0]{\begingroup \@sanitize@url \@href}%
\providecommand \@href[1]{\@@startlink{#1}\@@href}%
\providecommand \@@href[1]{\endgroup#1\@@endlink}%
\providecommand \@sanitize@url [0]{\catcode `\\12\catcode `\$12\catcode
  `\&12\catcode `\#12\catcode `\^12\catcode `\_12\catcode `\%12\relax}%
\providecommand \@@startlink[1]{}%
\providecommand \@@endlink[0]{}%
\providecommand \url  [0]{\begingroup\@sanitize@url \@url }%
\providecommand \@url [1]{\endgroup\@href {#1}{\urlprefix }}%
\providecommand \urlprefix  [0]{URL }%
\providecommand \Eprint [0]{\href }%
\providecommand \doibase [0]{http://dx.doi.org/}%
\providecommand \selectlanguage [0]{\@gobble}%
\providecommand \bibinfo  [0]{\@secondoftwo}%
\providecommand \bibfield  [0]{\@secondoftwo}%
\providecommand \translation [1]{[#1]}%
\providecommand \BibitemOpen [0]{}%
\providecommand \bibitemStop [0]{}%
\providecommand \bibitemNoStop [0]{.\EOS\space}%
\providecommand \EOS [0]{\spacefactor3000\relax}%
\providecommand \BibitemShut  [1]{\csname bibitem#1\endcsname}%
\let\auto@bib@innerbib\@empty
%</preamble>
\bigskip
%
\bibitem [{\citenamefont {Pardo}\ and\ \citenamefont
  {Pickett}(2009)}]{Pardo-Pickett_2009}%
  \BibitemOpen
  \bibfield  {author} {\bibinfo {author} {\bibfnamefont {V.}~\bibnamefont
  {Pardo}}\ and\ \bibinfo {author} {\bibfnamefont {W.~E.}\ \bibnamefont
  {Pickett}},\ }\href@noop {} {\bibfield  {journal} {\bibinfo  {journal} {Phys.
  Rev. Lett.}\ }\textbf {\bibinfo {volume} {102}},\ \bibinfo {pages} {166803}
  (\bibinfo {year} {2009})}\BibitemShut {NoStop}%
\bibitem [{\citenamefont {Banerjee}\ \emph {et~al.}(2009)\citenamefont
  {Banerjee}, \citenamefont {Singh}, \citenamefont {Pardo},\ and\ \citenamefont
  {Pickett}}]{Banerjee_2009}%
  \BibitemOpen
  \bibfield  {author} {\bibinfo {author} {\bibfnamefont {S.}~\bibnamefont
  {Banerjee}}, \bibinfo {author} {\bibfnamefont {R.}~\bibnamefont {Singh}},
  \bibinfo {author} {\bibfnamefont {V.}~\bibnamefont {Pardo}}, \ and\ \bibinfo
  {author} {\bibfnamefont {W.}~\bibnamefont {Pickett}},\ }\href@noop {}
  {\bibfield  {journal} {\bibinfo  {journal} {Phys. Rev. Lett.}\ }\textbf
  {\bibinfo {volume} {103}},\ \bibinfo {pages} {016402} (\bibinfo {year}
  {2009})}\BibitemShut {NoStop}%
\bibitem [{\citenamefont {Delplace}\ and\ \citenamefont
  {Montambaux}(2010)}]{Delplace-Montambaux_2010}%
  \BibitemOpen
  \bibfield  {author} {\bibinfo {author} {\bibfnamefont {P.}~\bibnamefont
  {Delplace}}\ and\ \bibinfo {author} {\bibfnamefont {G.}~\bibnamefont
  {Montambaux}},\ }\href@noop {} {\bibfield  {journal} {\bibinfo  {journal}
  {Phys. Rev. B}\ }\textbf {\bibinfo {volume} {82}},\ \bibinfo {pages} {035438}
  (\bibinfo {year} {2010})}\BibitemShut {NoStop}%
\bibitem [{\citenamefont {Saha}(2016)}]{Saha_2016}%
  \BibitemOpen
  \bibfield  {author} {\bibinfo {author} {\bibfnamefont {K.}~\bibnamefont
  {Saha}},\ }\href@noop {} {\bibfield  {journal} {\bibinfo  {journal} {Phys.
  Rev. B}\ }\textbf {\bibinfo {volume} {94}},\ \bibinfo {pages} {081103(R)}
  (\bibinfo {year} {2016})}\BibitemShut {NoStop}%
\bibitem [{\citenamefont {Banerjee}\ and\ \citenamefont
  {Narayan}()}]{Banerjee-Narayan_2020}%
  \BibitemOpen
  \bibfield  {author} {\bibinfo {author} {\bibfnamefont {A.}~\bibnamefont
  {Banerjee}}\ and\ \bibinfo {author} {\bibfnamefont {A.}~\bibnamefont
  {Narayan}},\ }\href@noop {} {\ }\bibinfo {note} {{a}rXiv:2001.11188
  [cond-mat.mes-hall] (2020)}\BibitemShut {NoStop}%
\bibitem [{\citenamefont {Kato}(1966)}]{Kato}%
  \BibitemOpen
  \bibfield  {author} {\bibinfo {author} {\bibfnamefont {T.}~\bibnamefont
  {Kato}},\ }\href@noop {} {\emph {\bibinfo {title} {Perturbation Theory for
  Linear Operators}}}\ (\bibinfo  {publisher} {Springer-Verlag},\ \bibinfo
  {address} {Berlin},\ \bibinfo {year} {1966})\BibitemShut {NoStop}%
\bibitem [{\citenamefont {Thaller}(1992)}]{Thaller}%
  \BibitemOpen
  \bibfield  {author} {\bibinfo {author} {\bibfnamefont {B.}~\bibnamefont
  {Thaller}},\ }\href@noop {} {\emph {\bibinfo {title} {The {D}irac
  equation}}}\ (\bibinfo  {publisher} {Springer-Verlag},\ \bibinfo {address}
  {Berlin Heidelberg},\ \bibinfo {year} {1992})\BibitemShut {NoStop}%
\bibitem [{\citenamefont {Bruneau}\ \emph {et~al.}()\citenamefont {Bruneau},
  \citenamefont {Jaksic}, \citenamefont {Last},\ and\ \citenamefont
  {Pillet}}]{Bruneau-Jaksic-Last-Pillet_2016}%
  \BibitemOpen
  \bibfield  {author} {\bibinfo {author} {\bibfnamefont {L.}~\bibnamefont
  {Bruneau}}, \bibinfo {author} {\bibfnamefont {V.}~\bibnamefont {Jaksic}},
  \bibinfo {author} {\bibfnamefont {Y.}~\bibnamefont {Last}}, \ and\ \bibinfo
  {author} {\bibfnamefont {C.-A.}\ \bibnamefont {Pillet}},\ }\href@noop {} {\
  }\bibinfo {note} {{a}rXiv:1602.01893 [math-ph] (2016)}\BibitemShut {NoStop}%
\bibitem [{\citenamefont {Davies}(1995)}]{Davies}%
  \BibitemOpen
  \bibfield  {author} {\bibinfo {author} {\bibfnamefont {E.~B.}\ \bibnamefont
  {Davies}},\ }\href@noop {} {\emph {\bibinfo {title} {Spectral Theory and
  Differential Operators}}}\ (\bibinfo  {publisher} {Camb. Univ Press},\
  \bibinfo {address} {Cambridge},\ \bibinfo {year} {1995})\BibitemShut
  {NoStop}%
\bibitem [{\citenamefont {Duclos}\ \emph {et~al.}(2001)\citenamefont {Duclos},
  \citenamefont {Exner},\ and\ \citenamefont {Krej\v{c}i\v{r}\'{\i}k}}]{DEK2}%
  \BibitemOpen
  \bibfield  {author} {\bibinfo {author} {\bibfnamefont {P.}~\bibnamefont
  {Duclos}}, \bibinfo {author} {\bibfnamefont {P.}~\bibnamefont {Exner}}, \
  and\ \bibinfo {author} {\bibfnamefont {D.}~\bibnamefont
  {Krej\v{c}i\v{r}\'{\i}k}},\ }\href@noop {} {\bibfield  {journal} {\bibinfo
  {journal} {Commun. Math. Phys.}\ }\textbf {\bibinfo {volume} {223}},\
  \bibinfo {pages} {13} (\bibinfo {year} {2001})}\BibitemShut {NoStop}%
\bibitem [{\citenamefont {Simon}(1976)}]{Si}%
  \BibitemOpen
  \bibfield  {author} {\bibinfo {author} {\bibfnamefont {B.}~\bibnamefont
  {Simon}},\ }\href@noop {} {\bibfield  {journal} {\bibinfo  {journal} {Ann.
  Phys.}\ }\textbf {\bibinfo {volume} {97}},\ \bibinfo {pages} {279} (\bibinfo
  {year} {1976})}\BibitemShut {NoStop}%
\end{thebibliography}
\end{document}